\documentclass[aps,prl, twocolumn,showpacs,longbibliography,10pt]{revtex4-1}

\usepackage{graphicx}
\usepackage{amsmath}
\usepackage[colorlinks=true,citecolor=blue,urlcolor=magenta]{hyperref}
\usepackage{tabularx}

\begin{document}

\title{Improved Limits on  Spin-Mass Interactions}

\author{Junyi Lee}
\email{junyilee@princeton.edu} 
\affiliation{Department of Physics, Princeton University}

\author{Attaallah Almasi}
\affiliation{Department of Physics, Princeton University}

\author{Michael Romalis}
\affiliation{Department of Physics, Princeton University}

\date{\today}
\pacs{0000}

\begin{abstract}
        Very light particles with  CP-violating  couplings to ordinary matter, such as axions or axion-like particles,    can mediate long-range forces between polarized and  unpolarized fermions. We describe a new experimental search  for such forces between unpolarized nucleons in two 250 kg Pb weights and polarized neutrons and electrons in a $^3$He-K co-magnetometer located about 15 cm away. We place improved constrains  on the products of scalar and pseudoscalar coupling constants,   $g^n_p g^N_s < 4.2\times10^{-30}$ and  $g^e_p g^N_s < 1.7\times10^{-30}$ (95\% CL) for axion-like particle masses less than $10^{-6}$ eV, which represents an order of magnitude improvement over the best previous neutron  laboratory limit. 
\end{abstract}

\maketitle

\section{Introduction}
        Spin-mass interactions between spin-polarized fermions and unpolarized fermions can arise in several ways. A particularly well-known and theoretically motivated mechanism arises from the exchange of an axion,  a light pseudo Nambu-Goldstone pseudoscalar boson \cite{Weinberg1978A,Wilczek1978Problem} associated with a spontaneously broken Peceei-Quinn U(1) symmetry which was originally postulated to solve the strong CP problem in QCD \cite{Peccei1977CP}. Its existence is of particular interest since it could potentially explain the unnaturally small level of CP violation in QCD \cite{Baker2006Improved,HgEDM} and also make up a significant fraction of the dark matter in the Universe \cite{Kim2010Axions}. In addition to the QCD axion, other light axion-like particles (ALPs) can arise from the breaking of additional U(1) symmetries, which for example, happens naturally in string theory \cite{Witten1984,Svrcek2006Axions,Arvanitaki2010String}.  

The spin-mass interaction mediated by a pseudoscalar particle involves a scalar and a pseudoscalar coupling, the strengths of which are parametrized by the dimensionless coupling constants $g_s$ and $g_p$ respectively. At tree level, this interaction yields the following "monopole-dipole" potential in the non-relativistic limit \cite{Moody1984New}: 
\begin{equation}
V(\mathbf{r})=\frac{\hbar^2 g_s g_p}{8\pi m_p}\hat{r}\cdot\hat{\sigma_p}\left(\frac{m_{\phi} c}{\hbar r}+\frac{1}{r^2}\right)e^{-r m_{\phi} c/\hbar},
\label{eq:SpinMassPotential}
\end{equation}
where  $m_p$ is the mass of the fermion  at the pseudoscalar vertex and $\hat{\sigma_p}$ is the normalized expectation value of its spin. $\hat{r}$ is the unit direction vector between the particles, $r$ is their separation distance, and $m_\phi$ is the mass of the axion. For the QCD axion, its mass and the strength of its fermion couplings are related to each other by the QCD anomaly, while for ALPs they can be considered as independent parameters. One can also construct models involving a very light spin-1 $Z'$ boson that would generate a long-range spin-mass force similar to Eq. (\ref{eq:SpinMassPotential}) if it has CP-violating couplings \cite{Fayet1996New,Dobrescu2006Spin}. 

Searches for  axions and axion-like particles have been conducted using a variety of methods. If they constitute  a substantial fraction of dark matter and have electromagnetic couplings, they can be detected by photon conversion in a strong magnetic field \cite{Sikivie}. Irrespective of dark matter content, axions have been searched for in emission from the Sun \cite{CAST}, photon-axion-photon conversion  ``light shinning through wall'' experiments \cite{Ballou}, and by looking for long-range forces other than gravitational and electromagnetic. The long-range force experiments can be  classified into mass-mass \cite{Adelberger}, spin-mass \cite{Bulatowicz2013Laboratory,Tullney2013Constraints,Youdin1996Limits,Venema1992Search,Wineland,Ni,QUAX} and spin-spin experiments \cite{Terrano2015Short,Vasilakis2009Limits}, which constrain different combinations of  $g_s$ and $g_p$ couplings. Astrophysical observations also place a number of constraints on axion parameters \cite{RaffeltRev}. Several  new axion search  methods have  been recently proposed and new experiments are being developed \cite{Graham,Geraci}.

Here we report the results of a new search for long-range forces using a  K-$^3$He co-magnetometer and a movable unpolarized source mass. The co-magnetometer measures the difference between interactions of electron spin in K and nuclear (primarily neutron) spin in $^3$He. Barring accidental cancellation, we set new limits on both neutron and electron couplings, improving laboratory  limits in the  axion mass range near $10^{-6}$ eV.
\section{Experimental Apparatus}
        The basic operating principles of a K-$^3$He co-magnetometer are described in more detail in \cite{Kornack2005Nuclear}. Briefly, the co-magnetometer uses overlapping ensembles of  spin-polarized K and $^3$He, which are strongly coupled via their Fermi-contact interaction during spin-exchange collisions \cite{Schaefer} when the resonant frequencies of K and $^3$He are matched \cite{Kornack2002Dynamics}. This coupling gives the co-magnetometer fast transient response and automatic cancellation of magnetic fields in all three directions.

\begin{figure} 
\center
\includegraphics[width=0.49\textwidth]{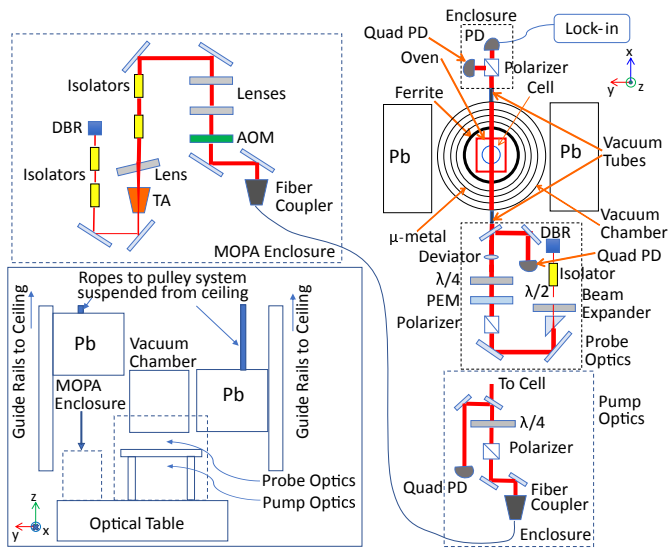}
\caption{(color online) Probe and pump optical setup in the x-y plane. Bottom left insert: Experimental schematic in the y-z plane.}
\label{Fig:Schematic}
\end{figure}

        K and $^3$He atoms are contained in a spherical glassblown cell made from GE180 glass with a diameter of 15 mm filled with 11 atm. of $^3$He (at room temperature), 20 Torr of N$_2$ for quenching and a droplet of K and $^{87}$Rb alkali metal.  The $^{87}$Rb is introduced in small quantities to enable hybrid optical pumping \cite{Babcock2003Hybrid}, which allows one to create a uniform alkali polarization in an optically-thick cell \cite{Romalis2010Hybrid}. At the cell operating temperature of 198$^\circ$C the density of Rb is measured to be $2\times10^{12}$ cm$^{-3}$ and the density of K is $1\times10^{14}$ cm$^{-3}$.  Under typical operating conditions  $^{87}$Rb is polarized to 70\% by optical pumping with 500 mW of 795 nm $\mathcal{D}$1 light. Rb-K spin-exchange collisions polarize K vapor to 20\% and K-$^3$He spin-exchange collisions polarize $^3$He to 2\%. A droplet of alkali metal is used to block the stem of the cell and we position the droplet with independent temperature control of the cell and stem in order to minimize spin-relaxation of $^3$He due to dipolar fields created by its own magnetization.

The cell is heated in a boron nitride oven with GaP windows \cite{GaP} using  thick film resistive heaters driven by AC currents at 120~kHz. The oven is placed  in  an inner ferrite shield and 3 outer $\mu$-metal magnetic shields \cite{ferrite}. The diameter of the cell is a compromise between spin-relaxation due to atomic diffusion to the walls, and the dimensions of the holes in the magnetic shields which limit magnetic shielding. In order to provide adequate thermal insulation and to reduce laser beam motion due to  air convection, the cell, oven,  and shields are placed in a vacuum chamber pumped to \textless100 mTorr.
 
The pump light is generated by a DBR laser amplified with a tapered amplifier (TA) and coupled into a single mode fiber. An accousto-optic modulator before the fiber allows  pump intensity feedback while keeping the amplifier injection current constant. This approach eliminates  laser pointing noise due to filamentation in the TA mode \cite{Goldberg1993Filament} and kinks  in the TA's output power at high injection current densities \cite{Fiebig2010Experimental}.

        The $x$-projection of K's spin, which constitutes the co-magnetometer's signal, is measured via optical rotation of a linearly polarized probe beam that is blue detuned from K's $\mathcal{D}1$ line. The 11 mW probe beam is supplied by a DBR laser tuned to $769.53$ nm and its plane of polarization is modulated with a photoelastic modulator and $\lambda/4$ plate combination. The optical rotation signal is then extracted by demodulating the raw signal from the photodiode with a lock-in amplifier (see Fig \ref{Fig:Schematic}).

During the experiment, the co-magnetometer is operated at the so-called compensation point where an applied B$_z$ field of about 3 mG approximately cancels the field that K atoms experience from  the magnetization of the $^3$He spins, which results in strong coupling between K and $^3$He spins. Residual magnetic fields inside the shields are canceled by coils within the ferrite shield and are adjusted by automated zeroing routines developed in \cite{Kornack2005Nuclear}. Under these near zero-field and high alkali density conditions, K is in a spin-exchange relaxation free regime \cite{Happer1973Spin,SERF} where the relaxation due to spin-exchange collisions is eliminated and high sensitivity is achieved. The signal of the co-magnetometer at the compensation point is approximately given by
\begin{equation}
S=\kappa\frac{P^e_z \gamma_e}{R^e_{tot}}\left(\beta^n_y-\beta^e_y+\frac{\Omega_y}{\gamma_n}\right),
\label{eq:Comag Signal}
\end{equation} 

where $\beta^{n}_y$ and $\beta^{e}_y$  are the anomalous magnetic field-like  couplings to  the nuclear  spin in $^3$He  and the electron spin in K. For ordinary magnetic fields  $\beta^{n}= \beta^{e}$, making the co-magnetometer insensitive to magnetic fields to first order.  $\Omega_y$ is the angular non-inertial rotation rate of the apparatus about the $y$-axis. It represents an example of a non-magnetic coupling to spin that does not cancel in the co-magnetometer.  $\gamma_{e}$ is the gyromagnetic ratio of the free electron while $\gamma_n$ is the gyromagnetic ratio of $^3$He. $P^e_z$ is the polarization of the K atoms and $R^e_{tot}$ is their total spin relaxation rate, and $\kappa$ is a  gain factor. The combination $\kappa P^e_z \gamma_e/R^e_{tot}$ serves as the calibration constant of the co-magnetometer signal in magnetic field units. It  can be measured using field modulation techniques by, for example, measuring the co-magnetometer's response to a slow modulation of the B$_x$ field \cite{Brown2010New}. We verified the calibration independently by detecting a signal induced by a slow rotation $\Omega_y$ of the optical table, which was also measured with a tilt sensor.

To generate the anomalous spin-mass interaction we use two stacks of Pb bricks, with overall dimensions of $30$~cm$\times\,36$~cm$\times\,20$ cm located along the y-axis of the co-magnetometer (see Fig \ref{Fig:Schematic}). The face of the blocks is 13.6~cm from the center of the cell. The positions of the blocks are periodically reversed  so that one of them is close to the cell while the other is raised by 50 cm.  This allows us to reverse the sign of the energy shift due to Eq.~\eqref{eq:SpinMassPotential} to search for a correlation between the position of the masses and the co-magnetometer's signal. The two Pb masses  are suspended from the ceiling using ropes in such a way that  the load on the ceiling is the same while they are at rest  in the two positions.   The weights are moved by a pulley system driven by a 4 kW servo motor connected to a reduction gear located close to the ceiling and far away from the co-magnetometer. The mechanical system is constructed using nylon pulleys, fiber ropes, and aluminum frame and fasteners to minimize  magnetic systematic effects.
 Furthermore, the dimensions of the system are designed so that the servo motor and reduction gear, which are magnetic, rotate an integer number of turns between the two positions and a $\mu$-metal shield is also placed over the motor and gear. The pump and probe optics are enclosed in nearly-airtight plexiglass boxes and the laser beams propagate toward the vacuum can through sealed glass tubes to eliminate air convection due to motion of the masses. Thermal insulation is placed around the vacuum chamber to minimize mK-level temperature changes correlated with the position of the weights that resulted in nm-level relative mechanical motion of the co-magnetometer components due to thermal expansion.
        
        Considerable effort was also taken to eliminate possible electromagnetic cross-talk between the power electronics of the motor and the co-magnetometer. A separate computer controls the motor motion through an optically-isolated interface. All equipment relating to the co-magnetometer is powered from a different circuit from the motor's equipment. Electromagnetic interference arising from the high frequency switching supply in the motor's servo amplifier was reduced by placing filters on its input and output. Additional filtering was also employed on sensitive co-magnetometer electronics and care was taken to eliminate any  ground loops there. Lastly, the servo motor was programmed to turn off after the end of each motion before the data used in the correlation analysis was taken. 

\section{Results and Discussion}

\begin{figure}
\centering
\input{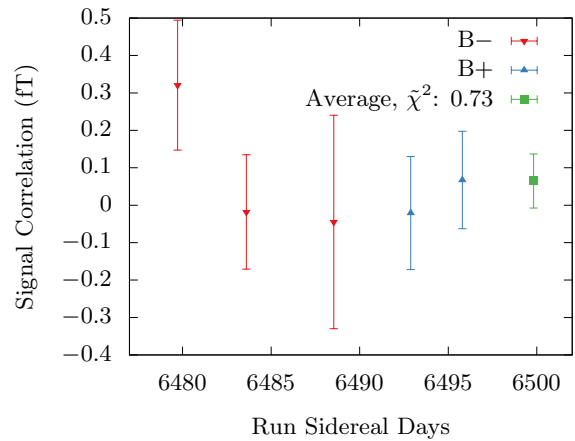}
\caption{Measured signal correlation. B$-$ runs are denoted with triangles pointing down while B$+$ runs are denoted with triangles pointing up. The weighted average is given by the square point. For 4 degrees of freedom, the $\tilde{\chi}^2$ is 0.73.}
\label{Fig:Signal Correlation Plot}
\end{figure}

        We collect data in records of 312 s after which the co-magnetometer undergoes an automated sequence of zeroing routines. During each record, the weights are moved 31 times so that the starting configuration of the weights alternate after each zeroing. The weights are moved over 5 sec and after a 3 sec wait, data from the next 2 sec is used in the analysis. A correlation "string" measurement \cite{Dress1977Search} is formed by taking an appropriate weighted  difference of the co-magnetometer's signal. At the end of an individual run, which typically lasts around 2 days, the strings are combined in a weighted average to give a correlation measurement for that run. In the K-$^3$He co-magnetometer the helicity of the pump light is fixed by the direction of the applied B$_z$ compensation field so that only two, B$+$ and B$-$, configurations are possible, which give opposite signs for possible spin-mass interactions. Figure \ref{Fig:Signal Correlation Plot} summarizes about 1.5 weeks of continuous data taking in both the B$+$ and B$-$ configurations. The reduced $\chi^2$ for these 5 runs is 0.73.

        Data from many other sensors are also collected in parallel and analyzed for correlations with the position of the weights. These include 4-quadrant photodetectors for laser position measurements, non-contact distance sensors to measure small motion, a 3-axis fluxgate magnetometer mounted outside the shields near the weights, and a 2-axis tilt meter with nrad resolution mounted on the optical table. 

Table \ref{Table:Result Summary} summarizes the results of these  measurements of possible  signal correlations. The sensors are calibrated by applying an appropriate excitation and measuring the main signal response and the sensor response. For example, the  fluxgate field measurements are multiplied by the measured combined shielding factor of the magnetic shields and the co-magnetometer compensation, which is $7 \times 10^8$ in the $x$-direction at the frequency pertinent to the experiment; other directions have higher suppressions.

\begin{table}[!htb]
\centering
\begin{tabular*}{0.45\textwidth}{@{\extracolsep{\fill}} cc}
\hline\hline
Sensor & Calibrated Correlation (aT) \\
\hline
Pump position  X & $ 2 \pm 4 $ \\ 
Pump position Y & $ -12 \pm 7 $ \\ 
Pump Intensity  & $ 2 \pm 5 $ \\ 
Fluxgate B$_x$ & $ -0.24 \pm 0.02 $ \\ 
Fluxgate B$_y$ & $ 0.03 \pm 0.01 $ \\ 
Fluxgate B$_z$ & $ -0.19 \pm 0.03 $ \\ 
Oven heater output & $ -1 \pm 1 $ \\ 
Vacuum can position  & $ -13 \pm 13 $ \\ 
Pump glass tube position & $ 2 \pm 2 $ \\ 
Probe position X & $ -12 \pm 9 $ \\ 
Probe position  Y & $ -3 \pm 6 $ \\ 
Probe Intensity & $ -7 \pm 6 $ \\ 
Rotation $\Omega_x$ & $ 1 \pm 1 $ \\ 
Rotation $\Omega_y$ & $ 0.1 \pm 0.7 $ \\ 
\hline
Total & $ -41\pm 20 $ \\
\hline\hline
\\
Signal (B$+$) & $ -5 \pm 100 $ \\
Signal (B$-$) & $ 70 \pm 100 $ \\
\hline
Signal & $ 32 \pm 70 $ \\
\hline\hline
\end{tabular*}
\caption{Summary of measured correlations in the signal and other calibrated sensors.}
\label{Table:Result Summary}
\end{table}

        We expect that systematic effects due to the  sensor correlations are independent and their uncertainties can  be combined in quadrature to provide an estimate of the overall systematic uncertainty.  Since most correlations are not statistically significant,  we do not correct the signal by the net sum of all measured correlations. Accordingly, we quote as the final result:
\begin{equation*}
\beta^{n}-\beta^{e}<(32 \pm 70_{\text{stat}} \pm 20_{\text{syst}}) \text{ aT,}
\end{equation*}
which gives an upper limit $| \beta^{e,n}|<155$ aT at 95\% C.L.
        
\begin{figure}
\centering
\input{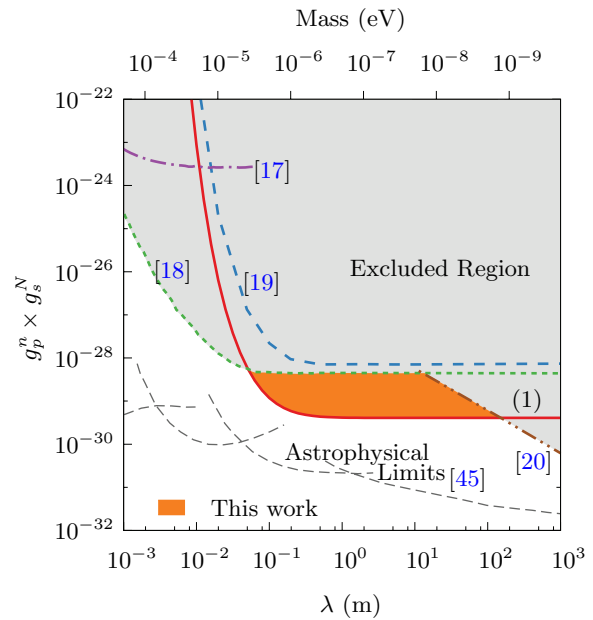}
\caption{(color online) Constraints (95\% CL) on $g^n_p g^N_s$ for the neutron. The solid red line (1) is from this work. Limits in \cite{Youdin1996Limits,Bulatowicz2013Laboratory} were multiplied by 2 to obtain a 95\% CL. \cite{Venema1992Search} was extrapolated by integrating over the Earth's mass density for heavier axions. \cite{Raffelt2012Limits} uses astrophysical constraints to limit $g^n_p$}
\label{Fig:Neutron Exclusion Plot}
\end{figure}

\begin{figure}[!ht]
\centering
\input{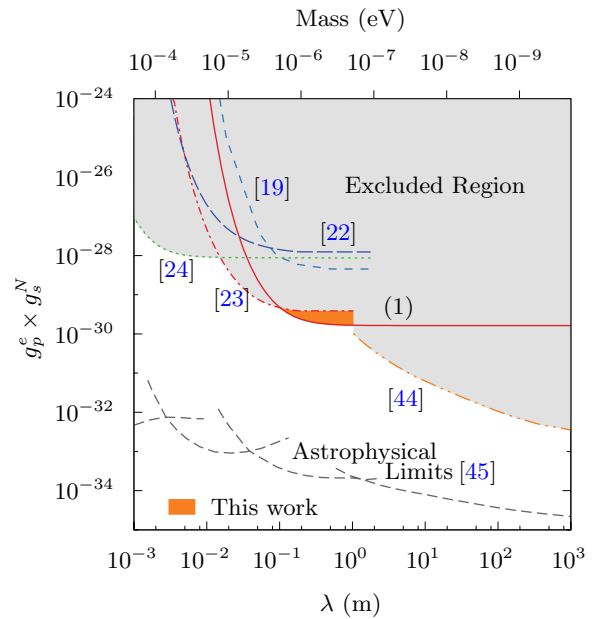}
\caption{(color online) Constraints (95\% CL) on $g^e_p g^N_s$ for the electron. The solid red line (1) is from this work.  Limits in \cite{Youdin1996Limits} were multiplied by 2 to obtain a 95\% CL. \cite{Raffelt2012Limits} uses astrophysical observations.}
\label{Fig:Electron Exclusion Plot}
\end{figure}

	As is typical, we assume that the scalar coupling to unpolarized fermions is the same for neutrons and protons and is zero for electrons in the unpolarized mass so that this coupling reduces to $g^N_s$, the axion's scalar coupling to nucleons. Similarly, we denote the axion's pseudoscalar coupling to polarized neutrons and electrons as $g^n_p$ and $g^e_p$ respectively. In order to derive the constrains on $g^n_p g^N_s$, we note that the neutron is 87\% polarized in $^3$He \cite{Friar1990Neutron,Ethier2013Comparative} and we therefore set $\mu_{^3\text{He}} \beta^n > g^n_p g^N_s  0.87A$, where $A$ is the numerical factor from the integration of Eq.~\eqref{eq:SpinMassPotential} over all the nucleons in the Pb masses. Similarly, we set $\mu_{B} \beta^e > g^e_p g^N_s A$ to derive the limits on $g^e_p g^N_s$. Figure \ref{Fig:Neutron Exclusion Plot} and \ref{Fig:Electron Exclusion Plot} show the respective constraints on $g^n_p g^N_s$ and $g^e_p g^N_s$. Our results set the most stringent laboratory limits for $m_\phi<3 \times 10^{-6}$eV. For much lower axion masses additional constrains can be obtained from experiments using the Earth as the source mass \cite{Venema1992Search,Heckel2008Preferred}.   

        In conclusion, we have placed new laboratory constraints on $g^n_p g^N_s$ and $g^e_p g^N_s$ using a K-$^3$He co-magnetometer and movable Pb masses. The limit on the neutron coupling is improved by an order of magnitude over two orders of magnitude in axion mass range. There is a significant potential for further improvement of the sensitivity since a similar co-magnetometer spin force experiment has achieved a sensitivity of 1 aT \cite{Vasilakis2009Limits}. The main challenge in the current experiment is due to mechanical motion of the masses instead of the electronic flip of the nuclear spins used in \cite{Vasilakis2009Limits}. This causes subtle mechanical effects due to temperature changes correlated with the positions of the masses. Further improvement will be possible by reducing the systematc sensitivity of the co-magnetometer to such effects. 

	This work was supported by NSF grant PHY-1404325.

%

\end{document}